\documentclass{aa}  
\usepackage{graphicx,natbib,ulem}
\usepackage{txfonts}
\begin{document}
  \titlerunning{Fluctuations and symmetry in the jets of SS\,433}
\authorrunning{Blundell, Bowler \& Schmidtobreick}
   \title{Fluctuations and symmetry in the speed and direction of the jets of SS\,433 on
   different timescales}

   \subtitle{}

   \author{Katherine M.\ Blundell \inst{1}
          \and
          Michael G.\ Bowler \inst{1}
          \and
          Linda Schmidtobreick \inst{2}
          }

   \offprints{K.M. Blundell \\   \email{katherine.blundell@physics.ox.ac.uk}}
   \institute{University of Oxford, Department of Physics, Keble Road,
              Oxford, OX1 3RH, UK \and European Southern
              Observatory, Vitacura, Alonso de Cordova, Santiago,
              Chile}
   \date{Received; accepted}

 
  \abstract
   {The Galactic microquasar SS\,433 launches oppositely-directed
  plasma jets at speeds approximately a quarter of the speed of light along an axis which precesses, tracing out a cone of polar angle 20 degrees. Both the speed and direction of launch of the bolides that comprise the jets exhibit small fluctuations. }
  {To present new results on variations in speed and direction of successive bolides over periods of one day or less and the range in speed and direction exhibited by material within individual bolides.  To present the expansion rate of bolides revealed by optical line broadening. To integrate these results with earlier data sensitive to fluctuations and symmetry over longer timescales.}
   {New high resolution spectra, taken with a 3.6-m telescope almost nightly over 0.4 of a precession cycle, are analysed in terms of fluctuations in the properties of the ejecta, by comparing the redshift data with predictions for small departures from the simple kinematic model. The variations of  redshifts of the jets within a day are studied, in addition to the longer term variations of the daily averaged properties.} 
  {(i) These data exhibit multiple ejections within most 24-hour
  periods and, throughout the duration of the observing campaign, the
  weighted means of the individual bolides, 
  in both the red jet and the blue jet, clearly exhibit the
  pronounced nodding known in this system.  (ii) We present further
  evidence for a 13-day periodicity in the jet speed, and show this
  cannot be dominated by Doppler shifts from orbital motion. (iii) We
  show the phase of this peak jet speed has shifted by a quarter of a cycle in
  the last quarter-century.  (iv) We show that the
  two jets ejected by SS\,433 are highly symmetric on timescales
  measured thus far. (v) We demonstrate that the
  anti-correlation between variations in direction and in speed
  is not an artifact of an assumption of symmetry. (vi) We show that a recently proposed mechanism (Begelman et al 2006) for varying the ejection speed and anti-correlating it with polar angle variations is ruled out. (vii) The speed of expansion of the
  plasma bolides in the jets is  approximately 0.0024\,$c$.   }
{These novel data carry a clear signature of speed variations. They  have a simple and natural interpretation in terms of both angular and speed fluctuations which are identical on average in the two jets. They complement archival optical data and recent radio imaging. }


   \maketitle
%

\section{Introduction}
Object number 433 in the catalogue of \citet{SS77} was
the first Galactic microquasar to be discovered \citep{Margon84}. It
is unusual in that it ejects, more or less continually, 
oppositely directed jets of gas (radiating particularly strongly in Balmer H$\alpha$) at a
mere 0.26\,$c$. The central engine is a member of a binary system with
an orbital period of 13.08 days (Crampton et al 1980, Margon et al 1980), but the natures of the
compact object and its companion are as yet not established. The jets are
remarkably well described, to first order, by a simple kinematic model
\citep{Milgrom79,FabianRees79,AbellMargon79}. In this model, the jets are ejected
oppositely to each other at an angle $\theta$ (approximately 20
degrees)
to an axis at angle $i$ (approximately 78 degrees) to our line-of-sight about which they
precess with a period of 162 days. The redshifts of the radiation
emitted by the bolides of plasma from the West jet ($z_{+}$, mostly
redshifted) and the East jet ($z_{-}$, mostly blueshifted) are
given, if the jets are symmetric, by:
\begin{equation}
\label{eq:redshift}
  z_{\pm} = -1 + \gamma[1 \pm \beta\sin{\theta}\sin{i}\cos{\phi} \pm
  \beta\cos{\theta}\cos{i}],
\end{equation}
where $\beta$ is the ejection speed and $\phi$ is the phase of the
precession cycle (see
http://www-astro.physics.ox.ac.uk/$\sim$kmb/ss433/).  We use the convention that $\phi$ is zero when the East jet is
maximally blueshifted.  In the
formulation of equation 1, $\beta$, $\theta$ and $\phi$ are common to the two
jets. Should the two jets not share these parameters, then
equation\,\ref{eq:redshift} is trivially modified by writing
$\beta_+$, $\beta_-$, $\theta_+$, $\theta_-$ and so on, so as to apply
to each jet individually. Superimposed on the simplest kinematic model
is the effect of nutation of (presumably) the accretion disc, known as
nodding \citep{Katzetal1982}.

Accumulated data on the Doppler shifts of the H$\alpha$ lines dating
back over 25 years have been reviewed by \citet{Eikenberryetal01} and
more recently by \citet{BlundellBowler05}. It has long been clear that
the redshifts exhibit fluctuations from the model predictions and that
there is a fair degree of symmetry in those
fluctuations.  (Indeed a striking episode was observed as early as 1978 --- see \citet{Margon79}.)   It was also long believed that these are due
to fluctuations in the angles $\theta$ and $\phi$ and owed little, if
anything, to changes in ejection speed of the
jets. \citet{BlundellBowler04} presented and analysed a very deep
radio image of the jets of SS\,433 corkscrewing across the sky. This
image constitutes an historical record of about two complete
precessional periods and shows very clearly that on a timescale of
10s of days the ejection speed of the two jets is variable with a rms
value of 0.014\,$c$ and furthermore that this variation is rather
highly symmetrical; if the East jet is abnormally fast the West is
also abnormally fast, by about the same change in $\beta$.

\citet{BlundellBowler05} subsequently showed that the archival
optical spectroscopy data also show speed variations of rms 0.013\,$c$, assuming
symmetry as revealed by the recent radio image, and also show an
anti-correlation between excursions in the speed $ \beta $ and in the
polar angle $\theta$ (also known as cone angle) 
in the sense that when faster jet bolides are
ejected, the polar angle is smaller. These newly discovered effects reproduced well the correlations between observed excursions $\Delta z_+$ and $\Delta
z_-$ \citep[figure 6 in][]{BlundellBowler05}.  These correlations are quite independent of
any assumption of symmetry, even though the model which explained them
was not. The analysis of the archival data revealed one further
novel feature, in that the sum of the redshifts of the two jets
(controlled entirely by speed in a purely symmetric model) exhibited a
periodicity of $\sim$13.08 days, the orbital period of the binary
system.

With these new results to hand, we were fortunate to
obtain nightly spectroscopic observations of SS\,433 extending, 
with few gaps, from August 2004
(Julian day 2453000+245.5) until SS\,433 became a daylight object
 (JD +321.5). The observations commenced close to
precessional phase zero (east jet maximally blueshifted) and continued
until precessional phase approximately $0.4 \times 2\pi$. With the
instrumentation available it was possible to detect and measure
accurately in a single spectrum up to as many as 10 separate Doppler
shifted H$\alpha$ lines in both jets and represent each by a Gaussian
profile. There is no known way, with the time-sampling available to
us, of pairing unambiguously multiple ejections in one jet with 
exact counterparts in the other jet and we
resorted to two ways around this limitation. The first (described in
Section 3) is to calculate for each jet in a single spectrum 
a weighted mean wavelength and
then pair the two means (which is unambiguous). The second (described
in Sections 4 \& 5) is to develop a method of studying fluctuations
within a given jet on a given day without using information about what
goes on in the other jet on the same day. Both methods yielded
valuable results and the second also casts new light on the
interpretation of the fluctuations observed in the archival data,
reported in \citet{BlundellBowler05}.  

\section{Observations}

Spectra were taken every 24 hours, each 
with approximately 5 minutes on-source time,
using the ESO 3.6-m New Technology Telescope on La Silla, Chile with the 
EMMI instrument, and its Grating \#\,6 together with a 0.5\,arcsec slit.
The resolution was 2.2\,\AA\ at 6000\AA\ and the
wavelength range covered from 5800 to 8700\,\AA. The raw
spectra were reduced with IRAF. SS\,433 is heavily
reddened \citep{Margon84} and near precession phase zero, bolides radiating
H$\alpha$ blueshifted to 5900 Angstroms were faint in comparison with their
counterparts at 7700 Angstroms. This was corrected to some extent by
supposing that the bright continuum spectrum from the almost
stationary parts of the system is flat. This is unlikely
to be adequate because even so corrected the blueshifted bolides were
not brighter than their redshifted counterparts. Finally, the
continuum itself dims during eclipse and corrections were made for
that effect. Fortunately the absolute intensities are of little
importance for the purposes of this paper.

Lines (5$\sigma$ or greater) above the continuum were fitted, using
Jochen Liske's software SPLOT, to gaussian profiles with three
parameters: central wavelength, standard deviation and height. Figure
1 shows a sample raw spectrum and the same spectrum after gaussian
representation.

\begin{figure}[htbp] 
   \centering
   \includegraphics[width=9cm]{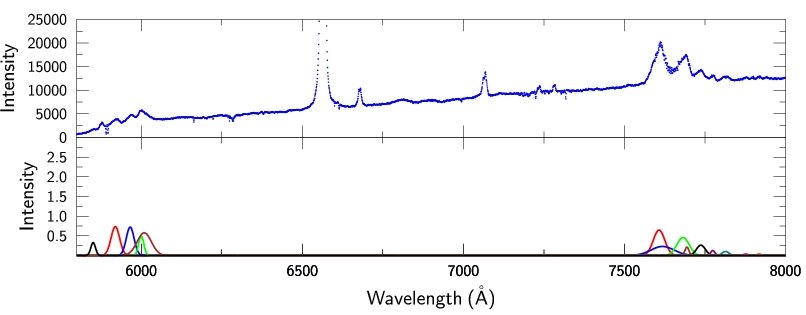} 
   \caption{Optical spectrum of SS\,433 on Julian day +260.5. The upper panel shows the spectrum after wavelength
   calibration and removal of the atmospheric telluric 
   features and, in
   addition to the ``moving'' Balmer H$\alpha$ lines, shows the
   ``stationary'' H$\alpha$ line, as well as HeI and CII lines.  The
   lower panel shows the gaussian fits to the moving H$\alpha$ lines
   after dividing through by a function fitted to the line-subtracted
   continuum background.  }
   \label{fig:spec17}
\end{figure}

Stationary lines were removed from this analysis and moving lines
identified as He or probable Paschen series hydrogen lines were also
deleted from the present study. 
The remaining lines were taken as Balmer H$\alpha$ associated with
the jet activity. These are shown day by day in Fig.\,2, where each
vertical mark has a height given by the logarithm of the product of
peak height and standard deviation of the fitted gaussian.                                                                                     The multiple ejections most days are striking and we refer to each such ejection as a bolide.
It is very clear that the scatter in measured wavelengths of bolides
within a given day is substantially greater in the West jet than in
the East for precessional phase near zero. It is also the case that
the widths of the individual lines are similarly greater in the West than in the East, although
that is not displayed here; see Fig.\,9 in Section 5. The nodding motion is visible
for the East jet, but because of the spread in the West jet the
nodding is not as easy to see there, but is clearly seen in the
plot of weighted means shown in Fig.\,3.

\begin{figure}[htbp] 
   \centering
   \includegraphics[width=9cm]{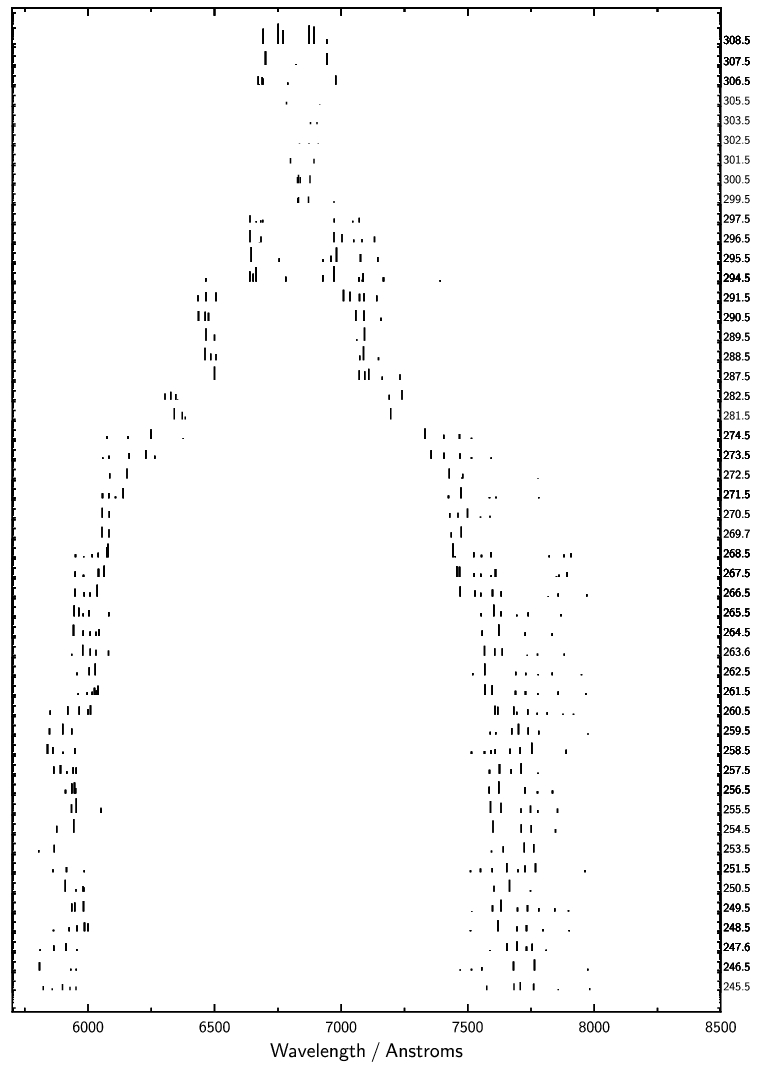} 
   \caption{The central wavelengths of ejected bolides are displayed day by day. Each horizontal series is from a single spectrum; Julian Day increases vertically. The height of each bar is the area of the fitted gaussian, on a logarithmic scale. The scatter in central wavelengths is much greater in the right-hand (West) jet than the left-hand (East) jet and nodding is apparent, especially clearly in the left-hand (East) jet.}
   \label{fig:ideagram}
\end{figure}

\section{Properties revealed by averaging within a day}

For each jet in each spectrum we calculated the mean
wavelength of the bolide complex, that is, the mean of the wavelengths of the individual bolides weighted by their respective areas. 
Each spectrum was thus represented by a single pair of
wavelengths as is the case for most archival data, some of which are of low
spectral resolution. The redshift history of these means is shown in
Fig.\,3. The nodding is very clear in both jets and mirror symmetry is
pronounced until after day 274. After day 285, both jets drift first in
the direction of decreasing redshift and then increasing redshift,
which could be (and probably is) due to decreasing and then increasing speed over durations of many days of the kind imaged by  \citet{BlundellBowler04}.

\begin{figure}[htbp] 
   \centering
   \includegraphics[width=9cm]{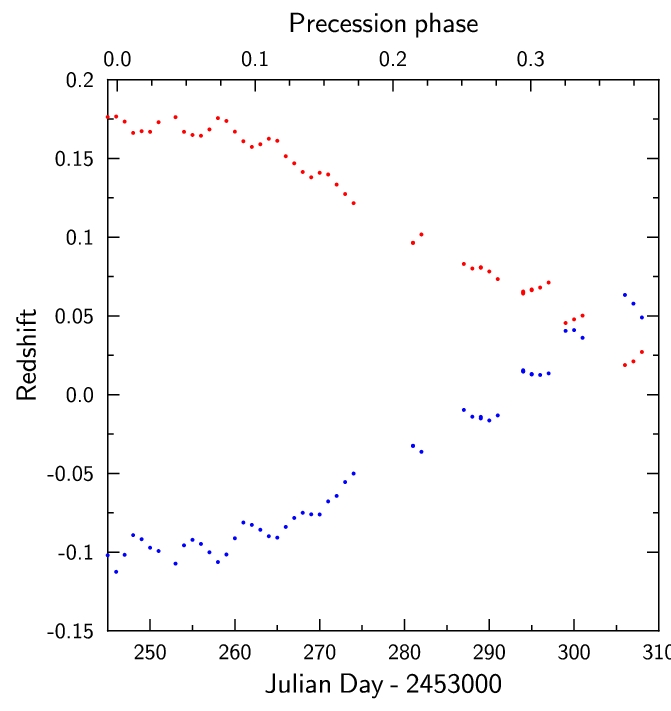} 
   \caption{The daily weighted mean redshifts of the ejecta in each jet as a function of time. The nodding is very clear in both jets and after day 285 both drift first to the blue and then to the red. }
   \label{fig:meanredshift}
\end{figure}

Under the assumption that the variation in jet parameters is perfectly
symmetric, the ejection speed can be extracted from the sum of the two
redshifts \citep{BlundellBowler05} (see also Marshall et al 2002).  These derived speeds are shown as
a function of time in Fig.\,4, where indeed the sustained higher
speeds beyond day 294 are apparent. There is a trace of an
approximately 13-day period in these speeds. To investigate how robust
this is, we took the spectra up to and including day 274 (before the
later long term drifts set in) and folded over 13.08 days, dividing
the orbital phase into 5 bins. The data and fitted curve are shown in
Fig.\,5 (lower panel). The signal is significant and the parameters are:

\begin{itemize}
\item Mean speed $0.260 \pm 0.002\,c$ 
\item Amplitude of speed variation  $0.006 \pm 0.001\,c$
\item Orbital phase of maximum speed $0.55 \pm 0.03 \times 2\pi$ (using the orbital
phase convention of \citet{Goranskiietal98}). 
\end{itemize}

\begin{figure}[htbp] 
   \centering
   \includegraphics[width=9cm]{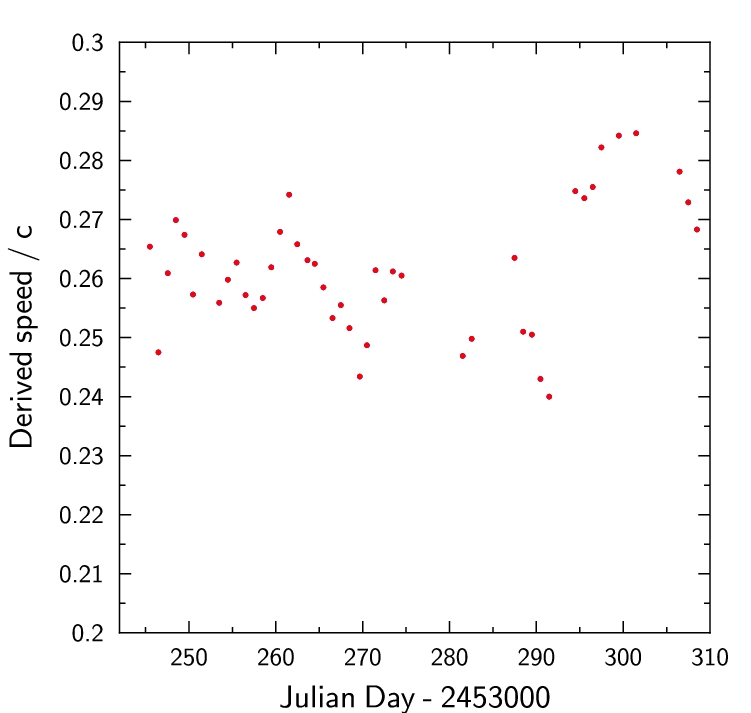} 
   \caption{If the mean redshifts in each jet are produced by identical variations of speed and angle, the common speed can be calculated from the sum of the redshifts. This figure displays the speeds so calculated. Speeds are abnormally high after day 294 and a 13-day periodicity is readily discernable.}
   \label{fig:speedday}
\end{figure}

The mean speed and the amplitude of speed variation are in excellent
agreement with those obtained from archival optical data
\citep{BlundellBowler05}.  However, the phase is considerably
different. The analysis of these new data thus drew our attention to
the orbital phase of peak velocity, a parameter we had not previously
investigated. We therefore refitted the archival data (Collins' data
set) and were careful to use the exact Goranskii ephemeris. The
results (see Fig.\,5, upper panel) were:

\begin{itemize}
\item Mean speed $0.2581 \pm 0.0005\,c$
\item Amplitude of speed variation $0.0063 \pm 0.0010$\,c
\item Phase of maximum speed $0.230 \pm 0.025 \times 2\pi$   
\end{itemize}

The Goranskii ephemeris is tied down by well-defined primary eclipses
observed over the period 1978 -- 1982 which dominates the archival
data. The visible effects of the primary eclipses on the continuum
background of our own data, presented in this paper, establish the
validity of the Goranskii ephemeris to an accuracy of better than a day for
our observations near the end of 2004. There is no doubt that the phase
of the oscillation in jet speed has shifted over a quarter of a cycle
between about 1980 and 2004.

\begin{figure}[htbp] 
   \centering
   \includegraphics[width=9cm]{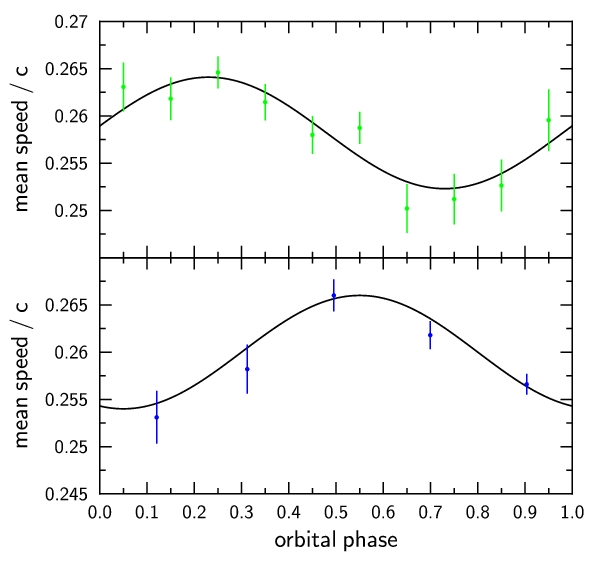} 
   \caption{ The figure shows the variation of jet speed, calculated from the sum of red shifts, folded over orbital phase bins, for (upper panel) the archival data and (lower panel) our 2004 data. Both sets show an orbital periodicity of the same amplitude, but the phases are different. The curves correspond to the fitted parameters listed in the text.}
   \label{fig:speedvorbitalphase}
\end{figure}

This 13-day periodicity occurs in the sum of $z_+$ and $z_-$, which
is a maximum at orbital phase 0.23 in the archival data (dominated by
1978 -- 1982) and 0.55 in the present data (observed 2004). If
interpreted as a variation in ejection speed, the ejection speed had
maxima at orbital phases 0.23 and 0.55 respectively. There are, however,
alternative explanations for a periodicity in the sum of the
redshifts. Perhaps the most obvious is simple Doppler shifting as a result of the orbital
motion of the compact object. 
If this were the explanation the orbital speed
would have to be in excess of 400\,km s$^{-1}$
\citep{BlundellBowler05} but in fact this explanation is ruled out by
the phases: at an orbital phase of 0.25, the compact object 
emerged from behind the companion star about a quarter of a 
revolution previously and so is approaching the
observer; $z_+ + z_-$ should be a minimum rather than the observed
maximum. The phase of the peak speed would also be fixed 
with respect to the orbital phase. 

         A further possible explanation for a 13-day
periodicity in the sum of $z_+$ and $z_-$ also has problems
with the phase difference between the archival and our 2004
data. Suppose that rather than being ejected anti-parallel, the two
jets are each deflected from the common jet axis of the kinematic
model by about 1 degree, always in a direction towards the
companion. That could do it for the data in this
paper, where the sum has a maximum at orbital phase $\sim$0.5. However, the plane containing the two jets could
not follow the companion perfectly because of the different phase observed 25 years ago. The
data are not sufficient to reveal whether the phase advances steadily
or can wander, but the existence of side bands having periods of 12.58 and 13.37 days
 in the archival data \citep[][fig.\ 3]{BlundellBowler05} suggests the latter. 
 
     Finally, there is the possibility that the ejection speed itself really does vary with a period of 13.08 days, perhaps as a result of Roche lobe overflow with a slightly eccentric orbit. If so, that orbit must have experienced a periastron advance of about 90 degrees in 25 years. Whatever the mechanism may be, the observation of
a 13-day periodicity in the sum of $z_+$ and $z_-$ may be revealing an
important property of the machinery whereby bulk plasma is accelerated
to one quarter of the speed of light.

\section{Fluctuations in redshift within a single jet}

Variation of the wavelength of
the emitted radiation due to thermal or bulk motion in a bolide rest
frame can be handled as described by eqns.(2-11) in this section, but for the case of spherical symmetry in the bolide rest frame
the observed line widths can also be attributed to a spread in proper
wavelength.  The observed line widths are too great for thermal
broadening since they would imply temperatures greatly in excess of $10^4$\,K.
 

Fluctuations in the wavelength emitted from a given jet
might be due to variation of speed within that jet, of polar angle
$\theta$ or of azimuthal angle $\phi$. (We discount the possibility of
the angle of inclination of the precession axis, $i$ in
equation\,\ref{eq:redshift}, varying.)   Differentiating equation\,\ref{eq:redshift} we have
\begin{eqnarray}
            \Delta z_+  &=&  F_+(\beta,\theta,\phi) \Delta \beta_+   +
            G(\beta,\theta,\phi) \Delta \theta_+ +
            H(\beta,\theta,\phi) \Delta \phi_+ \label{eq:deltazedplus}\\
            \Delta z_-  &=&  F_-(\beta,\theta,\phi) \Delta \beta_-   -
            G(\beta,\theta,\phi) \Delta \theta_-  -
            H(\beta,\theta,\phi) \Delta \phi_- \label{eq:deltazedminus}
\end{eqnarray}
where the same partial differentials $G$ and $H$ occur for both jets, but
because of the transverse Doppler effect $F_+$ and $F_-$ are different. It
is particularly important that when $\phi$ is approximately zero $F_+$ is much
larger than $F_-$.   The quantities $F{\pm}$, $G$ and $H$ are explicitly:
\begin{eqnarray}
F{\pm} &=& \beta\gamma^{3} {\pm} \gamma^{3} f_1\\
G &=&\beta\gamma f_2\\
H&=&\beta\gamma f_3
\end{eqnarray}
where the angular functions are given by
\begin{eqnarray}
f_1 &=& \sin \theta \sin i \cos \phi + \cos \theta \cos i \\
f_2 &=& \cos \theta \sin i \cos \phi - \sin \theta \cos i \\
f_3 &=& -\sin \theta \sin i \sin \phi.
\end{eqnarray}

If equations\,\ref{eq:deltazedplus} and \ref{eq:deltazedminus} are
each squared and we perform some appropriate average over the redshift
excursions for given (small range of) $\phi$, the observed $\langle
\Delta z_+^2\rangle$ are then parametrised in terms of 6 quantities
$\langle \Delta \beta_+^2\rangle$, $\langle \Delta \theta_+^2\rangle$,
$\langle \Delta \theta_+ \Delta \beta_+\rangle$ and so on. Each of
these six quantities is multiplied by a known function of the
precessional phase angle $\phi$. The explicit dependences on $\phi$ are 
given in eqns 10 and 11 below  and the
dependence of terms multiplying $\langle \Delta \beta^2\rangle$ and
$\langle \Delta \beta \Delta \theta\rangle$ is shown in
Fig.\,6. Fig.\,6 (upper panel) is in fact a different representation of figure\,1
in \citet{KatzPiran82}.

\begin{figure}[htbp] 
   \centering
   \includegraphics[width=9cm]{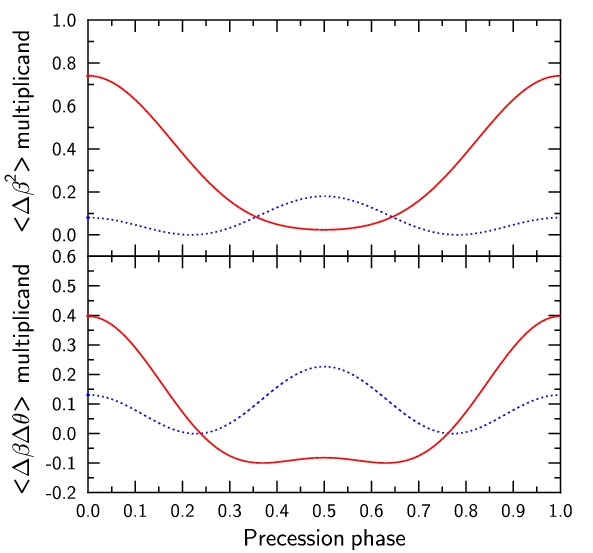} 
   \caption{The dependences on precessional phase of the terms in eqns. 10 and 11 which are multiplied by $\langle\Delta\beta^2\rangle$ (upper panel) and $\langle\Delta\beta\Delta\theta\rangle$ (lower panel). The East jet ($-$) is indicated in (broken) blue and the West (+) in (solid) red.  }
   \label{fig:wigglex}
\end{figure}

Then $\langle \Delta z_+^{2}\rangle$ is parametrised by
\begin{eqnarray}
\langle \Delta z_+^{2}\rangle &=& ( \beta \gamma^{3} + \gamma^{3}
f_1) ^{2}  \langle \Delta\beta_+^{2}\rangle \\ \nonumber
&+& (\beta \gamma f_2)^{2} \langle \Delta\theta_+^{2}\rangle   \\ \nonumber                                                                                                            
&+& (\beta \gamma f_3)^{2} \langle\Delta\phi_+^{2}\rangle \\ \nonumber     
&+& 2 (\beta \gamma^{3} + \gamma^{3} f_1) (\beta \gamma f_2)\langle \Delta \beta_+ 
\Delta\theta_+\rangle \\ \nonumber
&+& 2 (\beta \gamma^{3} + \gamma^{3}  f_1) (\beta \gamma f_3)\langle \Delta \beta_+ 
\Delta \phi_+\rangle \\ \nonumber
&+& 2 (\beta \gamma f_2) (\beta \gamma f_3)\langle \Delta \theta_+ \Delta \phi_+\rangle,
\end{eqnarray}
and $\langle \Delta z_-^{2}\rangle$ is parametrised by
\begin{eqnarray}
\langle \Delta z_-^{2}\rangle &=& (\beta \gamma^{3} - \gamma^{3} f_1)^{2} \langle\Delta
\beta_-^{2}\rangle  \\ \nonumber
&+& (\beta \gamma f_2)^{2} \langle\Delta\theta_-^{2}\rangle  + (\beta \gamma f_3)^{2} 
\langle\Delta\phi_-^{2}\rangle \\ \nonumber
&-& 2 (\beta \gamma^{3} - \gamma^{3} f_1) (\beta \gamma f_2)\langle\Delta\beta_-\Delta
\theta_-\rangle  \\ \nonumber
&-&  2 (\beta \gamma^{3} - \gamma^{3}  f_1) (\beta \gamma f_3)\langle\Delta\beta_-\Delta
\phi_-\rangle \\ \nonumber
&+& 2 (\beta \gamma f_2) (\beta \gamma f_3)\langle\Delta\theta_-\Delta\phi_-\rangle.
\end{eqnarray}

Thus given enough data of sufficient precision it would be possible to
determine all 6 quantities $\langle \Delta \beta_+^2\rangle$ etc for the
West jet and independently determine the analogous 6 quantities for
the East jet. No assumption of symmetry would then be required and if
the average $\langle \Delta \beta_+^2\rangle$ did actually differ
from $\langle
\Delta \beta_-^2\rangle$ that would be revealed by the data.

Unfortunately data sufficiently extensive and of such quality are not
available. It is however possible to determine
parameters such as $\langle \Delta \beta^2\rangle$ taken to be common
to the two jets on average only, without the extreme assumption of
symmetry in every ejection event (which we nonetheless believe to be a
very good approximation). We first apply this treatment to the
archival data and show that it leads to essentially identical results
as the assumption of maximal symmetry (variations in the parameters common to both jets) employed in \citet{BlundellBowler05}. In figure 6 of that paper it was shown that
the description of fluctuations in the jets extracted assuming
instantaneous symmetry also gave a very good description of the plot
of observed $\Delta z_-$ versus $\Delta z_+$ i.e.\ 
numbers extracted from the data
without any assumption of symmetry of any kind.  An unexpected result of \citet{BlundellBowler05} was a strong anti-correlation between
$\Delta\beta$ and $\Delta\theta$. It was pointed out (quite correctly)
by H.\ Marshall (private communication to KMB) that if speed variations are
asymmetric  then an assumption of symmetry would introduce a
specious anti-correlation. This of course does not mean that an anti-correlation is specious: if symmetry is true the anti-correlation is real.

In fact the $\phi$ dependence of $\Delta
z_+^2$ and $\Delta z_-^2$ requires both fluctuations in speed and also
separately an anti-correlation between $\Delta \beta$ and $\Delta
\theta$, if it be admitted only that the parameters averaged over many years are the same for both jets (that is, $\langle \Delta
\beta_+^2\rangle$ = $\langle \Delta \beta _-^2\rangle$ and so on for the other five pairs of parameters). Then subtracting equation 11 from equation 10, the difference of 
squared redshifts averaged over precessional phase is given by

\begin{eqnarray}
\langle \Delta z_+^2 \rangle - \langle \Delta z_-^2\rangle &=&  \\ \nonumber 
4 \beta \gamma^6 \cos\theta \cos i \langle \Delta \beta^2\rangle
- 4 \beta^2 \gamma^4 \sin\theta \cos i \langle \Delta \beta
\Delta \theta \rangle &=&  \\ \nonumber
0.252 \langle \Delta \beta^2\rangle - 0.022  \langle \Delta \beta \Delta\theta \rangle.  
\end{eqnarray}

Under these assumptions 
the difference between $\Delta z_+^2$ and
$\Delta z_-^2$ is given only by the terms in $\langle \Delta
\beta^2\rangle$ and in $\langle \Delta \beta \Delta \theta\rangle$
and the latter term is very small.  Fig.\,7 shows the measured difference in $\Delta z_+^2$ and
$\Delta z_-^2$ as a function of $\phi$ from Eikenberry's archival data set;
similar results are obtained with the data of Collins. 
The difference in $\Delta z_+^2$ and $\Delta z_-^2$ 
averaged over all $\phi$ is positive and has value $0.34 \pm0.08\times
10^{-4}$ (for red alone  --- i.e.\ $\Delta z_+^2$ --- the value is $1.22 \pm0.09\times 10^{-4} $)
and the difference between the red and blue jets is largest for
precessional phase 0.5.  Because the term in 
$\langle \Delta \beta \Delta \theta\rangle$ averaged   
over precessional phase is close to zero, the observed
difference requires a root mean square speed fluctuation of about
0.01\,$c$, regardless of the presence or otherwise of any
correlation between $\Delta\beta$ and $\Delta\theta$. 

 Then comparison of Fig.\,7 with Fig.\,6  
 shows that to produce the
 largest difference at precessional phase 0.5 rather than  zero
 absolutely requires an anti-correlation between $\Delta \beta$ and
 $\Delta \theta$ of about the size reported in
 \citet{BlundellBowler05}. The curve superimposed on Fig.\,7 is for
 $\langle \Delta \beta^2\rangle = 1.67 \times 10^{-4}$ and for
 $\langle \Delta \beta \Delta \theta\rangle = -3.8 \times
 10^{-4}$. It is also the case that $\langle \Delta z_+^2\rangle$ is
 larger near a phase of 0.5 than for phase close to zero, implying
 that a substantial anti-correlation is necessary in the red jet at
 least.  The anti-correlation between $\Delta \theta$ and $\Delta
 \beta$ is thus shown to be no artifact of the assumption of
 instantaneous symmetry made in \citet{BlundellBowler05}. However,
  an argument that the two jets are long term
 different cannot be refuted by the available
 Doppler data alone, but we repudiate it on the grounds of 
 the recent radio image \citep{BlundellBowler04}.

\begin{figure}[htbp] 
   \centering
   \includegraphics[width=9cm]{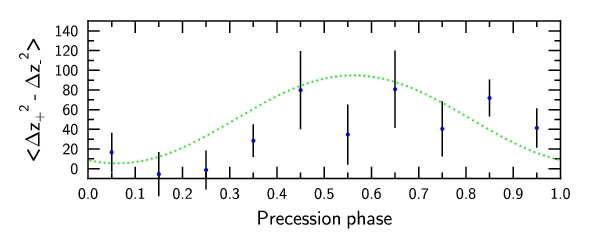} 
   \caption{The difference between the mean square fluctuations in redshift for the West jet and the East jet, as a function of precessional phase, averaged over many cycles in the archival data. The vertical scale is in units of $10^{-6}$. Averaged over all phases the red (West, +) jet has larger fluctuations than the blue (East, $-$). If the two jets are on average symmetric, this requires fluctuations in speed. The difference is small near precession phase zero and largest near phase 0.5; comparison with Fig.6 shows that this requires a strong anti-correlation between speed and direction. The curve is the prediction of \citet{BlundellBowler05}, rather than a fit. }
   \label{fig:demod}
\end{figure}

\begin{table*}[!t]
\caption{Comparison of short timescale fluctuations with long term
deviations.  Notes to Table. 1) The deviations in columns 2 and 3 
were fitted to data internal to jets on a given day, measuring
deviations from a local mean. The numbers in the last column represent
deviations from the simple kinematic model and are from the fits of
\citet{BlundellBowler05}, assuming instantaneous symmetry. These
numbers are also entirely consistent with analysis of those data
assuming only that many year averages are equivalent for the two
jets. 2) The last row gives the correlation
coefficient between $\Delta \beta$ and $\Delta \theta$, defined as
$\langle \Delta \beta \Delta \theta\rangle/ \sqrt{\langle\Delta
\beta^2\rangle\langle\Delta \theta^2\rangle}$.  }
\label{table:1}      
\centering          
\begin{tabular}{c l l l } 
\hline\hline       
fluctuating quantity & within bolides & among bolides & archival w.r.t kinematic model\\ 
\hline                    
$\langle \Delta \beta^2\rangle$ & $(0.08 \pm 0.01) \times 10^{-4}$ &  $(1.10 \pm 0.20) 
\times 10^{-4}$   
& $(1.67 \pm 0.18) \times 10^{-4}$ \\
rms $\Delta \beta$              &        0.0028           &                   0.01  &                   0.013 \\
& & & \\
$\langle \Delta \theta^2\rangle$ & $0.8 \pm 0.1 \times 10^{-4}$  &   $(1.5 \pm 0.4) \times 
10^{-4}$   
&  $(22.4\pm3.5) \times 10^{-4}$ \\ 
rms $\Delta \theta$  &    0.5 degrees      &              0.7 degrees &            2.71 degrees \\
& & & \\
$\langle \Delta \theta \Delta \beta\rangle$ & $(0\pm0.02) \times 10^{-4}$ & $(0\pm0.25) 
\times 10^{-4}$    
&  $(-3.81\pm 0.52) \times 10^{-4}$ \\
correlation coefficient &        $\sim 0$     & $\sim 0$     &    $\sim -0.62$  \\
\hline                  
\end{tabular}
\end{table*}

\section{Short timescale variations in the jets of SS\,433}

We have examined two measures of short timescale variations in the
jets of SS\,433, only possible because of the frequently-sampled high
quality spectra at our disposal. Both of these measures reveal
patterns which have simple and natural explanations if the average
properties of the two jets are the same. Fig.\,2 shows the pattern of
multiple resolved lines on any given night and one measure of short
timescale variation is the scatter of these lines in each jet in each spectrum.

For the bolides observed in each
jet on each night we calculated the area-weighted mean
square deviation in redshift of the individual lines from the weighted average of
all observed that night.  This quantity, the variance, is plotted in Fig.\,8 as a function of the
precessional phase.  The units are (dimensionless) redshift with a
factor of $10^{-6}$ extracted. There is a good deal of scatter from day
to day, but several features are clear. First, near precession phase zero the
West (red) has a much greater scatter in redshift than the East
(blue), by a factor of 4 or so. Second, the red and blue fluctuations 
converge as the precession phase advances towards $0.4
\times 2\pi$. Finally, both the red and the blue means decrease from 
phase zero  toward phase 0.3. Comparison of Fig.\,8 with Fig.\,6 (upper panel)
shows that these features are all characteristic of about equal rms fluctuations in each jet,
dominated by variation of the ejection speed.       

The second measure is, for each jet on each day, the area-weighted mean of the squares
of the standard deviations of the gaussians (in units of redshift with
$10^{-6}$ extracted) fitted to the individual bolides. This quantity, $\sigma^2$, is
plotted in Fig.\,9 as a function of precession phase. Here again the
mean square excursions are larger in the West (red) jet than in the
East (blue)  near phase of zero and the two converge as the phase
advances. In this instance the mean blue is rising as the mean red
falls. Again the role of speed fluctuations must be important but
their magnitude is much smaller than for Fig.\,8; pointing jitter
dominates in the East jet. In fact, this pattern is close to --- and
could be --- that from spherically symmetric expansion in
the rest frames of the individual bolides, on average the same in the
two jets.

\begin{figure}[htbp] 
   \centering
   \includegraphics[width=9cm]{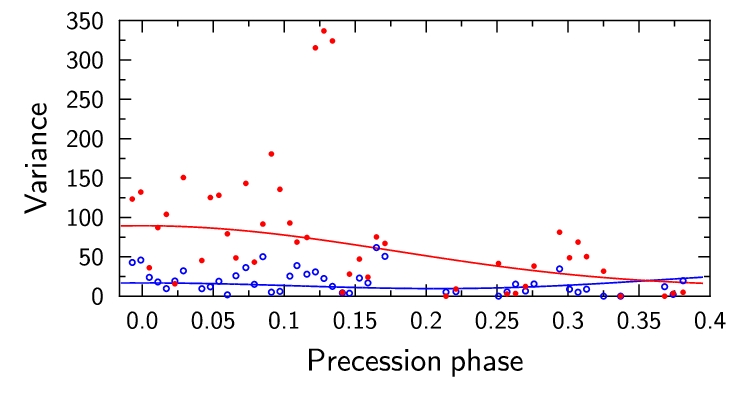} 
   \caption{The scatter of bolide redshifts, measured by the weighted variance discussed in the text, as a function of precession phase. The vertical scale is in units of $10^{-6}$. 
The West jet is indicated by red, solid symbols; the East jet by blue, open symbols. The curves are constructed from eqns. 10 and 11, with the parameters listed in Table 1.  The red and blue patterns in conjunction bear the signature of dominant fluctuations in speed; see Fig.\ 6.   }
   \label{fig:stats8}
\end{figure}

   These data shown in Figs\,8 and 9 are scattered in variance and $\sigma^{2}$ respectively, but do not extend beyond precessional phase of $0.4 \times
2\pi$; in order to quantify the contents of Figs.\,8 and 9 we made the
following assumptions. First, we assumed that the quantities $\langle
\Delta \beta^2\rangle$, $\langle \Delta \beta \Delta \theta\rangle$
and so on, the averages being taken over a few days (rather than over
many years as for the archival data) are common to both jets, as
suggested by the distinctive patterns in those figures. Second, we
assumed that the rms value of $\Delta \phi$ is given by the
rms value of $\Delta \theta$, divided by $\sin \theta$.  
This is the condition for ``pointing jitter''
\citep{KatzPiran82} and was satisfied in analysis of the archival data
in \citet{BlundellBowler05} whether or not the jets were required to
be anti-parallel.  We also assumed that the averages $\langle \Delta
\beta \Delta \phi\rangle$ and $\langle \Delta \theta \Delta
\phi\rangle$ could be neglected. The data in both jets were thus
represented in terms of three parameters $\langle \Delta
\beta^2\rangle$, $\langle \Delta \theta^2\rangle$ and the correlation
parameter $\langle \Delta \beta \Delta \theta\rangle$. The raw data-points
shown in Figs.\,8 and 9 were averaged within 10 bins of precessional
phase, the errors on the means taken from the variance of the entries
within a bin (but set very large in the rare cases where a bin
contained only two entries or less) and the $\chi^2$ minimised in
terms of those three fluctuation parameters. The resulting smooth curves
are shown in Figs.\,8 and 9 and the extracted parameters summarised in
Table 1 and compared with the parameters from the archival data. The
simple model assuming average symmetry of the two jets provides as
good a description of these data as anyone could desire. 

It is however the case that the features of Fig.\,8 could be explained by
fluctuations dominated by $\Delta \theta$ provided that such
fluctuations are systematically greater (by more than a factor of two)
in the West jet than in the East. Similarly, the features of Fig.\,9
could be explained by systematically different variations of $\Delta
\theta$ and $\Delta \phi$ in the two jets. However, such explanations are contrived and unnatural.

\begin{figure}[htbp] 
   \centering
   \includegraphics[width=9cm]{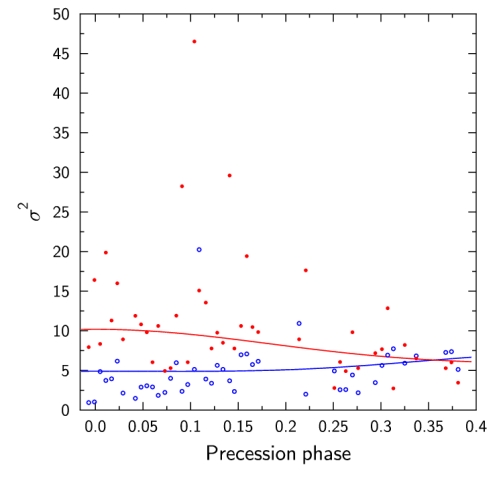} 
   \caption{The spread of the widths in redshift of the lines observed from individual bolides, as measured by the square of the standard deviation of the fitted gaussians, displayed as a function of precessional phase. The East jet is indicated by blue, open symbols and the West by red, solid symbols. The curves are constructed from the parameters listed in Table 1. The blue distribution is rising to meet the red; this corresponds to relatively small fluctuations in speed. If the two jets each contain bolides expanding spherically in the bolide rest frame, much the same pattern would be generated, in which case the magnitude of the red curve would be reduced by about a factor of 0.8. The vertical scale is in units of $10^{-6}$. }
   \label{fig:stats9}
\end{figure}

\section{Bulk motion and expansion of the plasma }

Material within an individual ejection (a single bolide or gaussian fit from our 2004 Chile data) on average has 
has an rms angular spread of about half a degree. There is a
significant spread in speed but much smaller than the difference in
speeds of separate bolides emitted within a day or so. This is hardly
surprising since it is required that bolides be resolved.  Over a day
or so the ejection speeds of successive individual bolides vary by more than
0.01\,$c$ and this variation is close to the rms deviation 
from the kinematic model in the archival data. However,
the observed spread of angular deviations within a day is a factor of 4 smaller. 
It would seem that it is much easier to eject different speeds within a day
than to twist the nozzles on the same timescale, perhaps not
surprising if the nozzles are embedded in a disc with enormous moment
of inertia. There is no significant anti-correlation between
$\Delta\beta$ and $\Delta\theta$ within a single bolide; nor between
bolides within a single day. In contrast, there is substantial (anti-) correlation in
deviations from the kinematic model, sensitive to much longer timescale
effects.

The data on the spread within individual bolides are consistent with a
rather simple picture. The curves shown in Fig.\,9 are for a combination
of speed fluctuations and pointing jitter within individual bolides. 
The corresponding numbers listed in Table 1 show longitudinal speed
variations of rms value 0.0028\,$c$ and rms transverse speed
variations of 0.0023\,$c$. This transverse speed is $\beta$ (0.26\,$c$) multiplied by the rms angular spread (the square root of $\langle\Delta\theta^2\rangle$) extracted from Fig.\,9. 
The curves for the two jets also converge
in just the same way that the redshifts themselves converge. Thus
these data suggest that the ``within bolide" characteristics are those of
optically thin fireballs expanding with spherical symmetry within
their own rest frames. If this model is imposed on these data the rms
speed of expansion is 0.0024$\,c$, greatly in excess of any plausible thermal speeds.

It is a pity that SS\,433 became a day-time object before a precessional
phase of even 0.4 was reached; completing half a precessional rotation
should have been much more informative because according to the model
fit the East (or left or blue) jet would have exhibited inter-bolide
variations exceeding those in the West (or right or red) jet in the
vicinity of precession phase 0.5. The consistency of the simple model (in which the two jets are equivalent at least on average and speed fluctuations are present) with
these new data is nonetheless compelling. 

\section{Conclusions}

The jets of SS\,433 exhibit fluctuations in direction and in speed on all
time scales measured thus far. The new data reported here reveal,  
for material within individual bolides, an rms angular spread of about 0.5\,degrees,
with rms (longitudinal) speed fluctuations of 0.0028\,$c$, rather like
spherical expansion of optically thin bolides in their own rest frames. 
Where resolved bolides are produced within
the same 24-hour period, the rms pointing jitter of those bolides is only a
little larger, 0.7\,degrees, but the spread of speeds is
about 0.01\,$c$.  There is no evidence that material within single bolides shows an anti-correlation between $\Delta\beta$ and  $\Delta\theta$, nor between bolides on short timescales of order one day collectively exhibit such anti-correlation.   

These results depend on assuming that the East and West jets are
equivalent averaged over periods of a few days or less, but do not
require instantaneous symmetry. The analysis of archival data in
\citet{BlundellBowler05} assumed instantaneous symmetry and that
analysis described well the archival data with rms angular deviations
from the kinematic model of 2.7\,degrees and rms speed fluctuations of
0.013\,$c$, with a correlation coefficient of $-0.62$.
 In this paper we have shown that these results are
also obtained assuming only that the two jets are equivalent when
averaging over the many year timescale of the archival data. There can
now be little doubt that the two jets are highly symmetric on
timescales longer than a day or two; the extent to which the jets are
symmetric on timescales of a day or less cannot be answered without
high precision sampling of the optical spectrum as frequently as every
few hours.

It is extremely interesting that the speed variations exhibited by
multiple bolides within a day (rms 0.01\,$c$) are pretty much
identical to the speed variations on much longer timescales
\citep{BlundellBowler04,BlundellBowler05} yet the associated angular
fluctuations are much smaller on timescales of order one day. It is also
interesting that any correlations between angular and speed
fluctuations are of much smaller magnitude on the shorter time
scales. These observations alone make it unlikely that angular tilts,
in for example the disc, control the perceived ejection speed
\citep{Begelmanetal06} but it might be that sustained higher speed
ejection is capable of reaming
out the nozzle in the disc and so decreasing the polar angle.

It should be emphasised that the physical origin of the anti-correlation 
between $\Delta \beta$ and $\Delta \theta$ in the long
term data 
(correlation coefficient = $-0.62$) is quite unknown  --Ð but
we have now demonstrated clearly that this result of \citet{BlundellBowler05}
is not an artifact
of the assumption of instantaneously symmetric ejection speeds as there employed. 

In a recent paper, \citet{Begelmanetal06} suggested that this anti-correlation might be
produced by the jets impacting on the walls of the core of the disc
and if they retained only that component of speed parallel to the walls. However, this
is inconsistent with the archival data and the 2004 data presented here, because the angular fluctuations
are not big enough to produce the associated speed fluctuations. In the model of \citet{Begelmanetal06} the emergent jet velocity is given by         
\begin{eqnarray}
  \beta = \beta_0\cos\theta
 \end{eqnarray}
 where $\beta_0$ is a constant speed before impact on the walls and hence 
 
 \begin{eqnarray}
   \Delta \beta = - \beta_0\sin\theta \Delta \theta
  \end{eqnarray}  
Then $\langle \Delta \beta^2\rangle$ would be approximately equal to 0.01 $\langle \Delta \theta^2\rangle$ and this relationship is completely inconsistent with Table 1.  In addition, if some trace of this mechanism were in
operation the nodding frequency would appear in the sum of $z_+$ and
$z_-$ but it does not \citep{BlundellBowler05}. The new observation of
negligible correlation on short time scales further rules out the explanation suggested by \citet{Begelmanetal06}.

The multiple ejections within a single day and their properties are
themselves of great interest, and we have shown that for the purposes
of pairing ejecta even the area-weighted mean redshifts contain a
great deal of information; they exhibit beautifully symmetric nodding
and also systematic drifts associated with longer-term speed
changes. We found that the interval from day 245 to 274 showed a clear
periodicity of 13 days in the sum of $z_+$ and $z_-$,
with maximum excursion in the 2004 data
when the compact object and the
companion both lie along the line of sight.  In the archival data
\citep{BlundellBowler05} the maximum excursion is a quarter of a cycle
earlier (dominated by observations of 25 years ago) but in both cases
the maximum excursion is in quite the wrong place for this periodicity
to be a Doppler effect due to the orbital speed of the compact
object. To obtain greater insight, it is necessary to mount a campaign of at least daily observations sustained over some
years.

\begin{acknowledgements}
The new observations reported in this paper were made possible by the
grant of Director's Discretionary Time on the 3.6-m New Technology
Telescope. We thank Jochen Liske for his excellent line-fitting
software, SPLOT. We are very grateful to the Leverhulme Trust whose
support has benefitted this work.  KMB thanks the Royal Society for a
University Research Fellowship.
\end{acknowledgements}

\end{document}